# Morphology and Photocatalytic Activity of Highly Oriented Mixed Phase Titanium Dioxide Thin Films

D.A.H. Hanaor [1]*, G. Triani [2], and C.C. Sorrell [1]

[1] University of New South Wales, School of Materials Science and Engineering, Sydney, NSW 2052, Australia

[2] Australian Nuclear Science and Technology Organisation, Institute of Materials and Engineering Sciences, Menai, NSW 2234, Australia

**Abstract**

Thin $TiO_2$ films on quartz substrates were prepared by spin coating of undoped and metal-ion-doped Sol-Gel precursors. These films were characterised by Scanning Electron Microscopy, Laser Raman Microspectroscopy, X-ray Diffraction and UV-Vis Transmission. The photocatalytic performances of the films were assessed by the photo-degradation of methylene-blue in aqueous solution under UV irradiation. Films exhibited a high degree of orientation and a thermal stabilization of the anatase phase as a result of substrate effects. In the absence of dopants, the rutile phase forms as parallel bands in the anatase which broaden as the transformation progresses. $TiO_2$ films doped or co-doped with transition metals exhibited the formation of rutile in segregated clusters at temperatures under ~800°C as a result of increased levels of oxygen vacancies. Photocatalytic activity of the films synthesised in this work was low as likely result of poor $TiO_2$ surface contact with dye molecules in solution. The presence of transition metal dopants appears detrimental to photocatalytic activity while the performance of mixed phase films was not observed to differ significantly from single phase material.

Keywords: $TiO_2$; Rutile; Anatase; Photocatalysis; Thin Films; Morphology

# 1. Introduction

$TiO_2$ photocatalysts have attracted a significant volume of research interest during the last 30 years owing to the various environmentally beneficial applications of such materials. While thin films of $TiO_2$ exhibit poorer surface area, and thus lower photocatalytic activity in comparison with suspended powders, the synthesis of durable thin film coatings of $TiO_2$ is attractive as such materials eliminate the requirement for catalyst recovery processes and can be applied to impart beneficial qualities to existing surfaces. The various applications of such $TiO_2$ photocatalytic coatings include

Self cleaning coatings [1-4]

Self sterilising (antimicrobial) coatings [5-7]

Water purification [8-10]

Photocatalysed reactions on $TiO_2$ surfaces take place through the generation of electron-hole pairs, known as excitons, by radiation exceeding band gap of the material (~3.0eV for rutile and ~3.2eV for anatase [11-13]) and the subsequent formation of adsorbed reactive species. The performance of a $TiO_2$ photocatalyst is limited by the relative rates of photo-generation of excitons, generally by UV irradiation, and their recombination. Both of these competing phenomena are influenced by the morphology and phase composition of the material. It has been reported that $TiO_2$ of a mixed anatase-rutile phase assemblage exhibits superior photocatalytic performance as a result of improved charge carrier separation through the trapping of conduction band electrons in the rutile phase [14-17]. Morphologies of smaller grain size are reported to show superior photo-activity as a result of a higher surface area and thus greater levels of adsorbed reactive species [18, 19].

While rutile is the more thermodynamically stable phase of $TiO_2$ at all pressures and temperatures [20-22], anatase is frequently the first phase formed in many synthesis routes due to a less constrained structural re-arrangement necessary to form this phase from an amorphous precursor and a lower surface energy in comparison with the rutile phase [23-25]. Upon heating in air, the phase transformation of anatase to rutile generally takes place around 650°C in the absence of dopants or impurities. Dopants influence the anatase to rutile phase transformation through a change in oxygen vacancy levels and the inhibition or promotion of the structural rearrangement involved in the transformation [26-29].

The performance of $TiO_2$ photocatalysts is often studied by the photo-degradation of sample pollutants in liquid, solid or gas phase. Such experiments are usually carried out under UV illumination although the use of natural sunlight is attractive due to the lower environmental impact involved. Methylene blue is frequently used to examine photocatalytic activity due to the ability to use spectrometric techniques to determine the concentration of the dye.

The present work examines the morphology, phase assemblage and photocatalytic activity of mixed phase thin-film $TiO_2$ photocatalysts. Although it has been reported that apart from the use of noble metals (Pt, Au and Ag) most metal doping is detrimental to photocatalytic activity of titania[30-32] , the present work included the use of metal dopants to examine the consequent effects of dopants on the phase assemblage of biphasic $TiO_2$ thin films and the prospect of increased photon absorption through inter-valence charge transfer between dopants of different valences.

# 2. Experimental Procedure

## 2.1. Materials

The preparation of samples involved the following chemicals. Isopropanol (>99% Univar) Titanium-tetra-iso-propoxide or TTIP (97% Sigma Aldrich), 2M solution of reagent grade Hydrochloric Acid in distilled water, and regent grade iron chloride ($FeCl_3$) and copper chloride ($CuCl_2$).both from Sigma Aldrich. 20x20x1 mm (100) quartz wafers were used as substrates. Quartz substrates were used rather than glass slides due to the increased thermal stability of quartz and due to the problems associated with increased diffusion of cations from glass [33, 34].

## 2.2. Preparation of Sol

Undoped films were made using a 50ml 0.5 M solution of TTIP in isopropanol prepared under magnetic stirring. For hydrolysis purposes 2M HCl was added to this solution in a quantity calculated to give a 2:1 molar ratio of $H_2O$:Ti. An increase in the viscosity of the solution was observed subsequent to this addition. A copper doped sol was prepared using the same method with the addition of copper chloride to give 5 at% Cu (Cu:Ti ratio 1:20), a co-doped sol was prepared with the additions of iron chloride and copper chloride both at 2.5 at% (Cu/Fe:Ti ratio 1:40).

### 2.3.Thin Film Synthesis

A Laurell WS-650Sz Spin coater was used to fabricate thin $TiO_2$ coatings on 20x20x1 mm quartz substrates. The substrate surfaces were covered with sol using a pipette followed by spinning at 1000RPM. This was repeated three times to increase film thickness. Films were fired in air in a muffle furnace at temperatures ranging from 600-1000 degrees for durations of 4 hours using a 5°/min heating rate. These thermal treatments imparted good film-substrate adhesion as thin films were resistant to rubbing off. Using a Micropack NanoCalc-2000 thin film measurement system in reflection mode with an integration time of 400 mS, films were determined to have thicknesses between 50 and 70 nm.

### 2.4. Analysis of films

An Invia Laser Raman Microspectrometer was used in conjunction with a 514 nm laser for the purpose of phase analysis with respect to microscopically visible features in the films. Overall phase analysis was carried out by glancing angle x-ray diffraction using a Phillips MRD system. Phase quantification was carried out using the method developed by Spurr and Myers [35]. This method is described by equation 1.

$$\frac{W_R}{W_A} = 1.22\frac{I_R}{I_A} - 0.028 \qquad (1)$$

Here $W_R$ and $W_A$ are respectively the weight fractions of rutile and anatase ($W_R$=1-$W_A$), $I_R$ and $I_A$ are respectively the intensity of the rutile (110) peak at 27. 35° 2θ and the anatse (101) peak at 25.18° 2θ.

Optical transmittance of the films prepared in this work was determined using a Lambda-35 UV-Vis Spectrophotometer in the wavelength range 300-800 nm.

Microstructure of films was examined using Hitachi s3500 and S900 scanning electron microscopes.

### 2.5. Photocatalytic activity

The decomposition of methylene blue (MB) was carried out in a cylindrical vessel containing 1 litre of MB with an initial molar concentration of 0.107 mMol/litre. The glass substrate was supported on a platform in the solution and the solution was irradiated with a 15 W UV lamp with an emission peak at 350 nm wavelength placed 100mm above the film. Irradiation was applied intermittently for alternating intervals of 15 minutes to avoid excessive heating of the solution. Experiments were carried out under magnetic stirring and air sparging. Optical absorbance of the methylene blue solutions were carried out after a cumulative 10 hours and 20 hours of irradiation had been achieved. MB concentrations were calculated using the measured absorbance peak at or around 663nm in conjunction with a prepared calibration procedure similar to analytical methods reported in the literature [36-38].

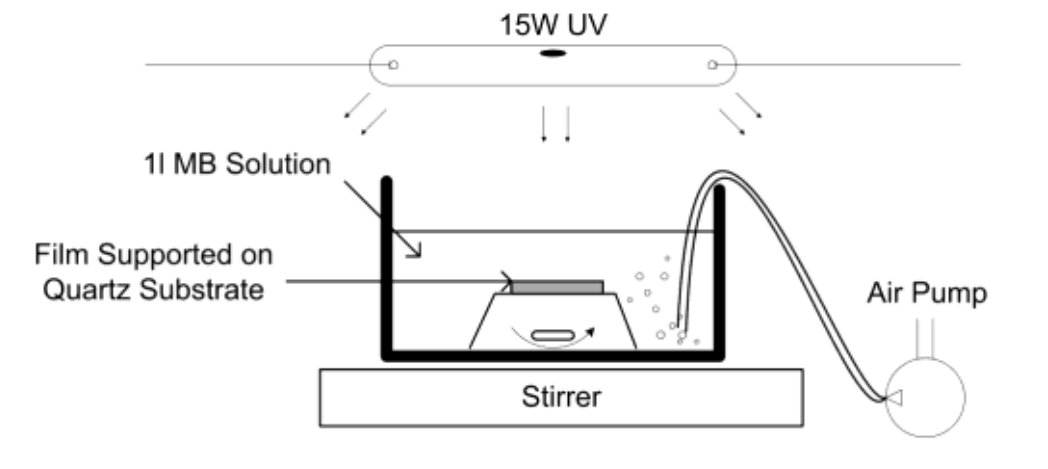

**Figure 1. Reactor for the photo-degradation of methylene blue by thin film $TiO_2$ photocatalysts**

## 3. Results

### 3.1. XRD

XRD patterns were gather from undoped, Cu doped and Cu/Fe co-doped films fired at different temperatures. The resultant XRD spectra are shown in figure 2. The dominance of the anatase (101) and rutile (110) peaks suggest a high degree of grain orientation. The Lotgering orientation factor was not calculated owing to the high noise to signal ratio in the XRD patterns gathered. Using the method of Spurr and Myers, these XRD spectra were interpreted to give anatase and rutile phase proportions which are plotted in Figure 3. The presence of phases other than anatase or rutile was considered negligible in these calculations and thus the sum of anatase and rutile fractions was assumed to be unity.

Undoped films showed a biphasic composition when fired at temperatures between 800°C and 1000°C. At 1100°C the transformation was complete and only the rutile phase was observed. In comparison with undoped films, the anatase to rutile transformation was promoted by Cu doping as evident in figures 2b and 3, showing a predominantly rutile phase after firing at 800°C. Cu/Fe co-doped films also exhibited enhanced anatase-rutile transformation showing complete transformation to the rutile phase after firing at 1000°C.

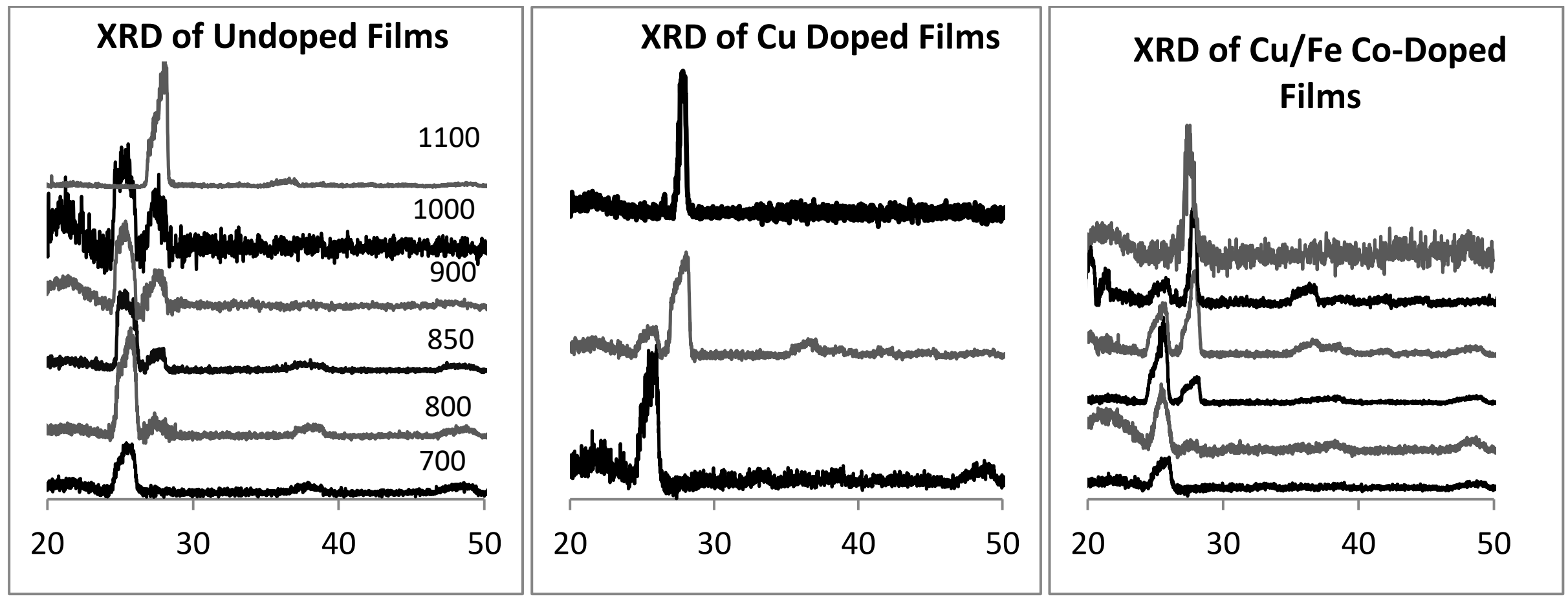

**Figure 2. Glancing angle XRD spectra of (a)Undoped $TiO_2$ films (b) Cu(ii) doped thin films and (c) Cu(ii)/Fe(iii) Co-Doped films fired at different temperatures**

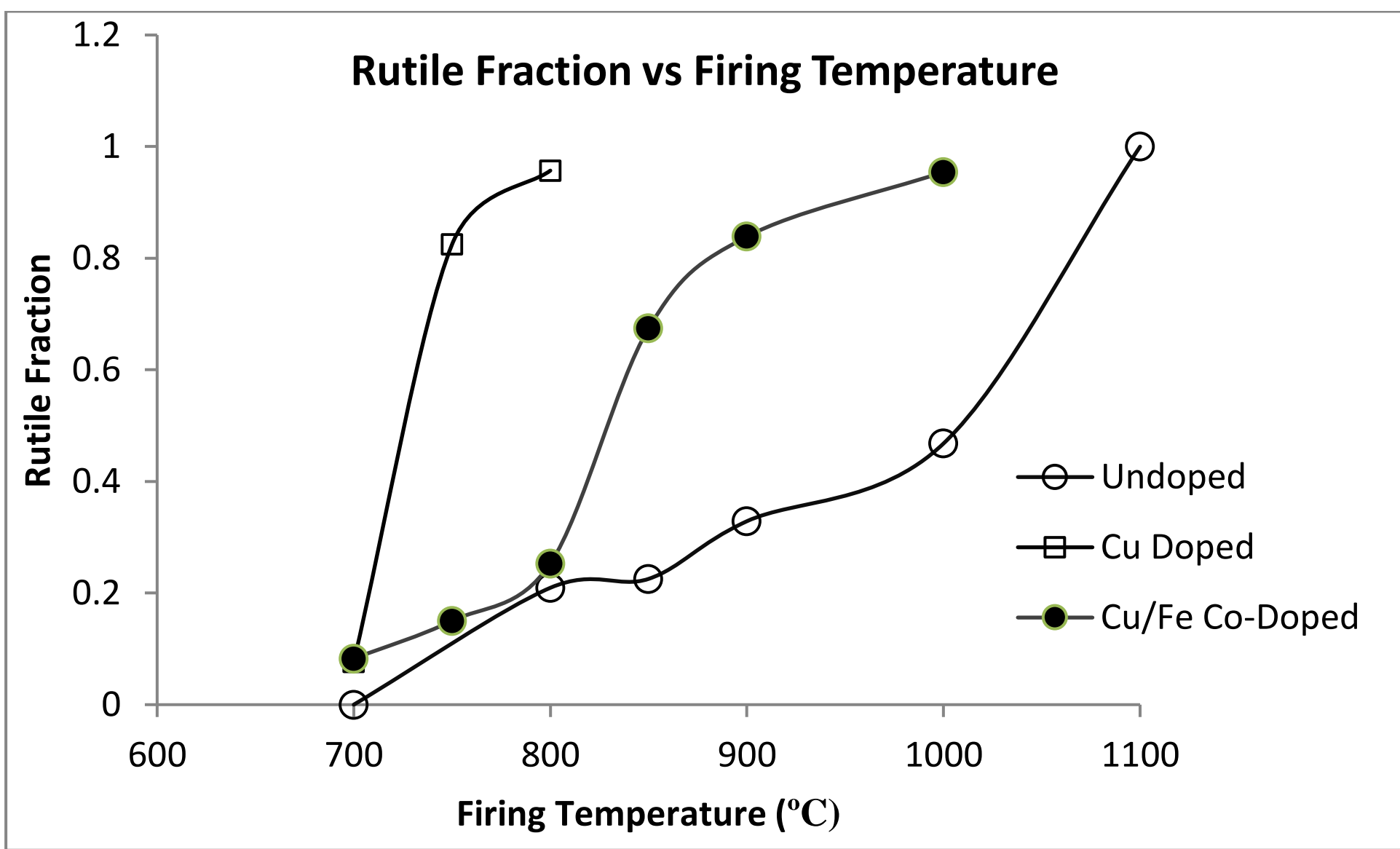

**Figure 3. Rutile content as a function of firing temperature for undoped, Cu(ii) doped and Cu(ii)/Fe(iii) Co-Doped $TiO_2$ thin films fired for 4h**

### 3.2.

### 3.3. Raman Microspectroscopy

The use of laser Raman microspectroscopy in conjunction with an optical microscope allows the differentiation of $TiO_2$ phases with respect to microscopically visible features. Figures 4-6 show micrographs of thin films alongside Raman spectra gathered from the regions of different appearance apparent in the images. It was found that undoped samples exhibit the formation of rutile as parallel bands in anatase (Figure 4) these bands showed a consistent orientation throughout the film. Rutile bands became wider and anatase bands more narrow as the phase transformation progressed. In contrast, doped films showed segregated clusters of rutile as spots of varying dimensions in the parent anatase (Figures 5 and 6). It should be noted that the dominance of the anatase peak at 141 $cm^{-1}$ does infer that this is the dominant phase rather this is a result of the interaction area of the laser and the particularly strong Raman shift of anatase at this wave-number.

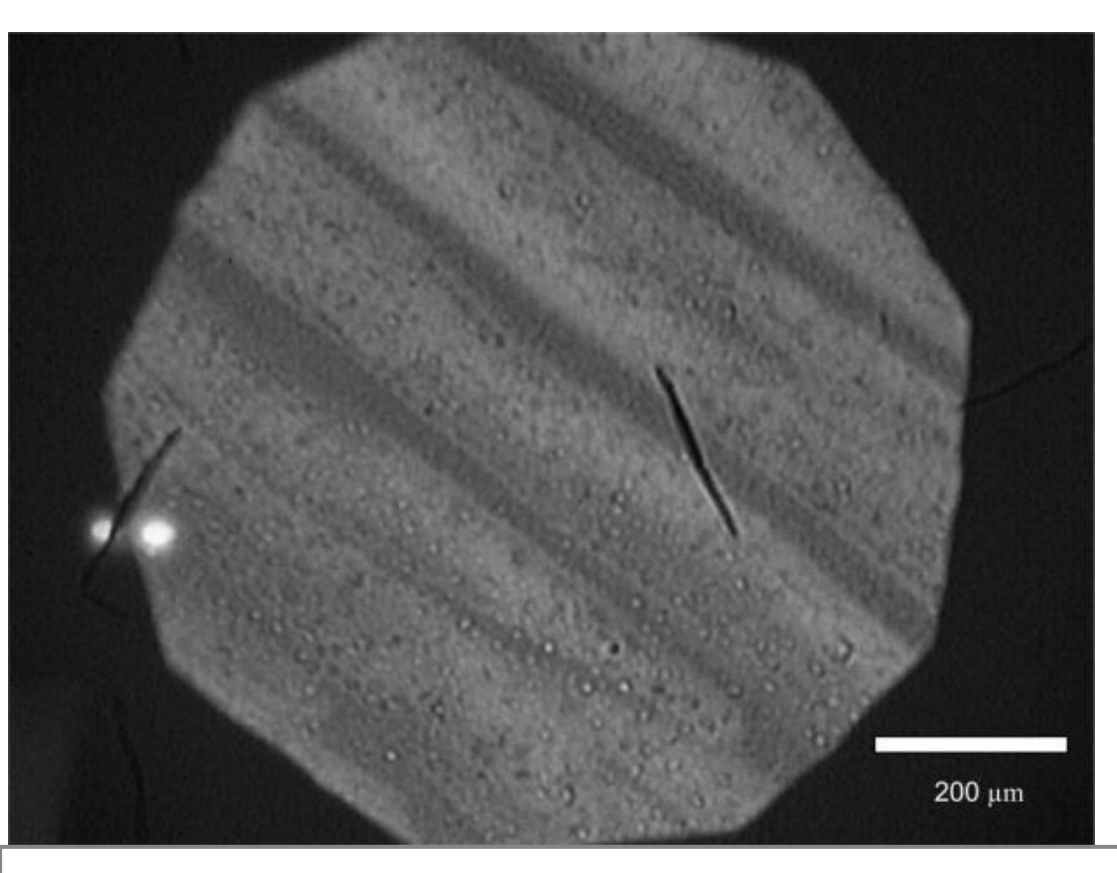

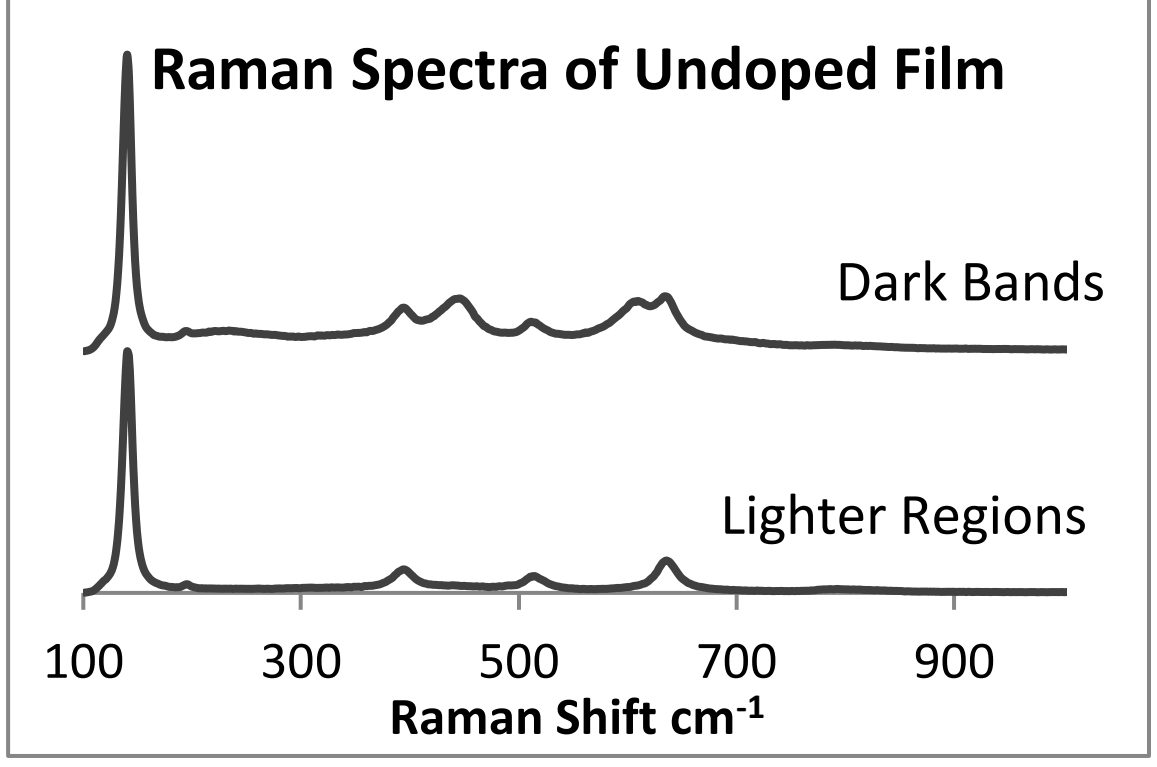

**Figure 4. (a) Optical micrograph of a 30% Rutile undoped $TiO_2$thin film (b) Raman spectra gathered from spots in darker and lighter bands**

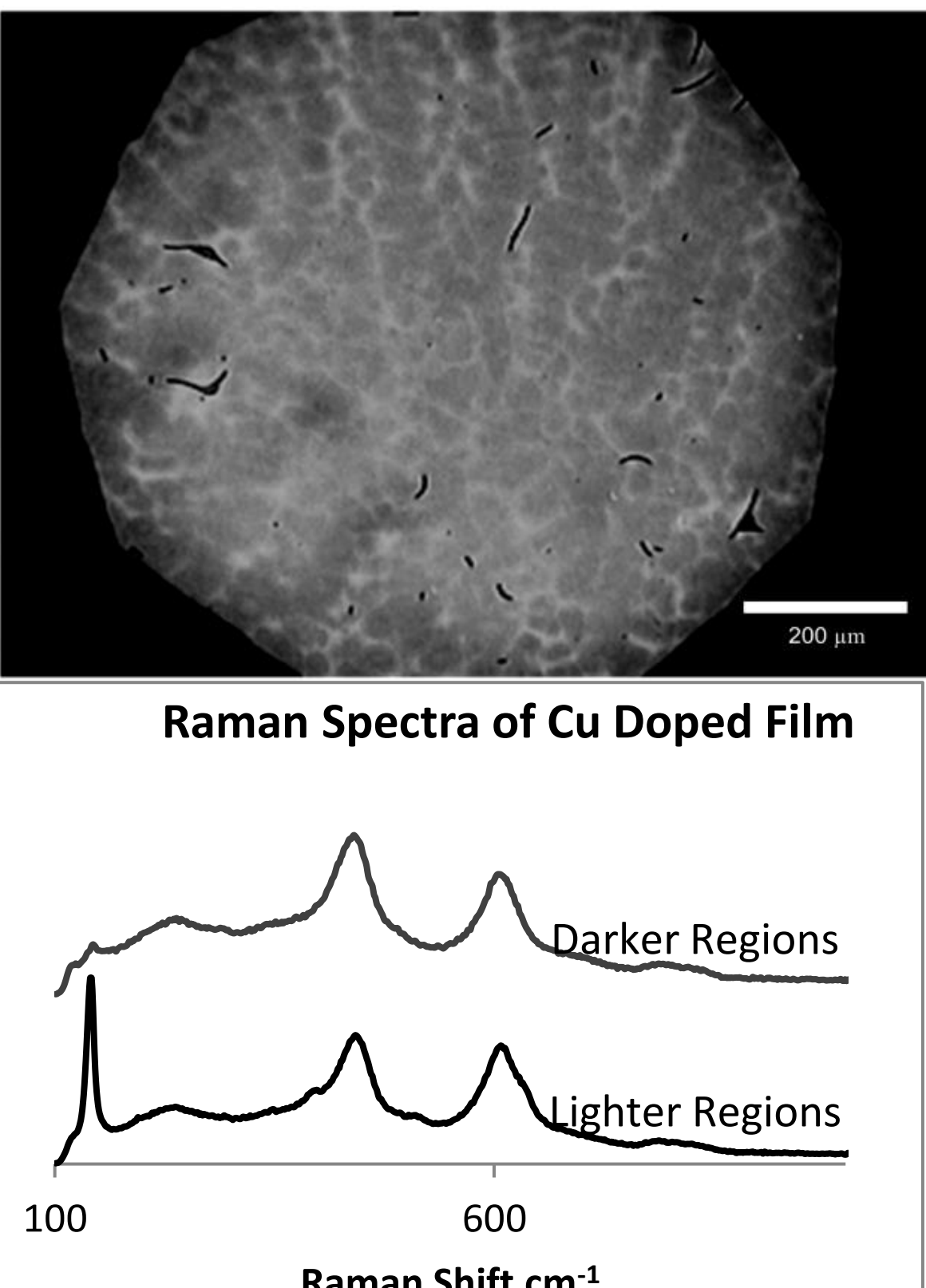

**Figure 5. (a) Optical micrograph of an 82% Rutile Cu(ii) Doped $TiO_2$thin film (b) Raman spectra gathered from dark spots and from lighter regions between spots.**

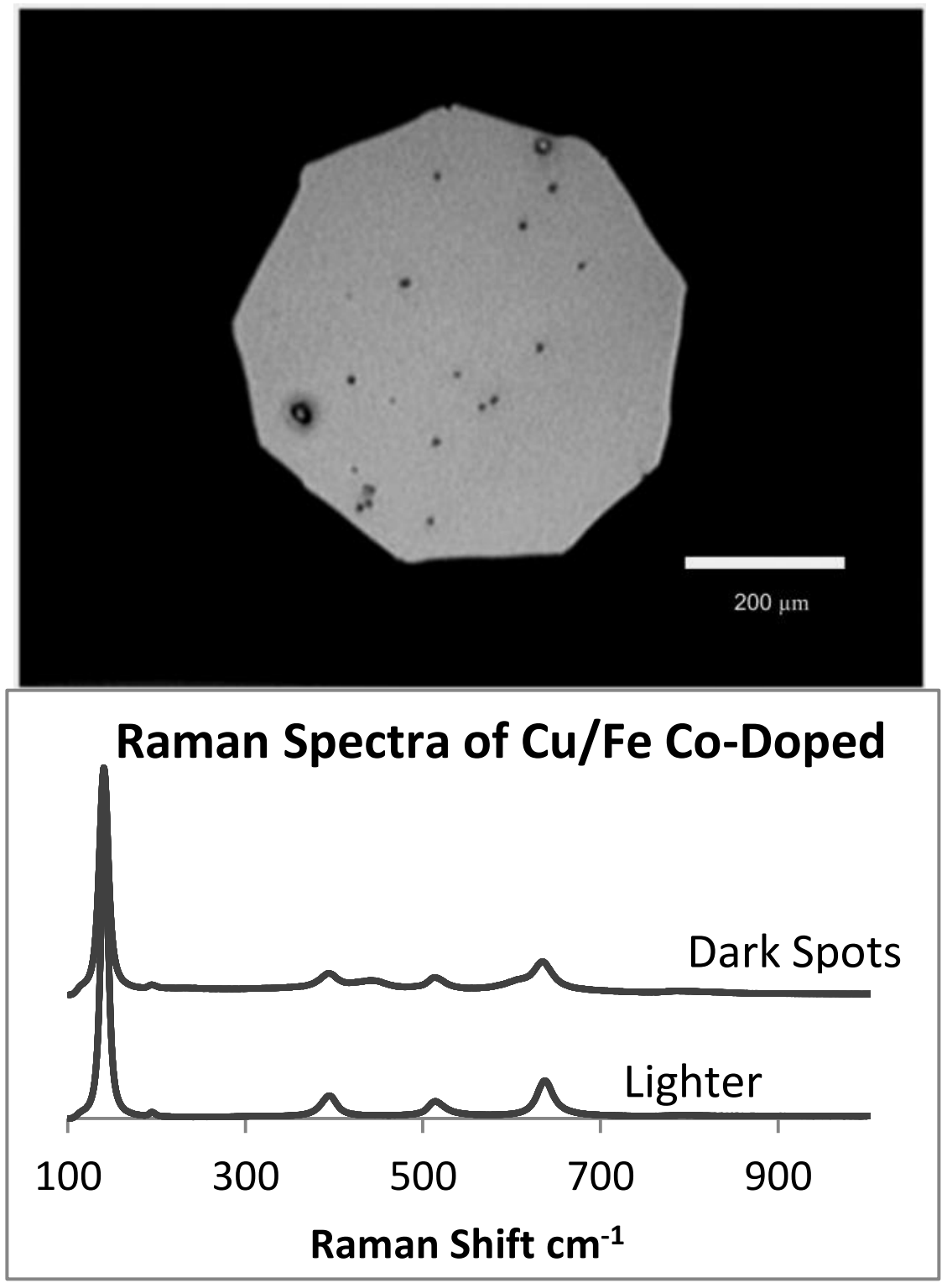

**Figure 6. (a) Optical micrograph of a 15% Rutile Cu(ii)/Fe(iii) Co-Doped $TiO_2$thin film (b) Raman**

**spectra gathered from dark spots and from lighter regions.**

## 3.4. SEM Analysis

Scanning electron microscope images of undoped films show an increase in grain size with the transformation to rutile (Figure 7). It was found that films formed in this work were fully dense, the anatase phase showed a grain size of approximately 40-50 and rutile exhibited larger grains of 200-400nm. Little to no porosity was observed however some shrinkage cracking was present. Grain morphology was similar in doped and co-doped anatase films, however less grain coarsening takes place in the transformation to rutile as evident from the absence of large consolidated grains in figure 8. The smaller crystallites visible in figure 8(b) are likely to be from the sputtered chromium coating applied to the sample prior to SEM analysis.

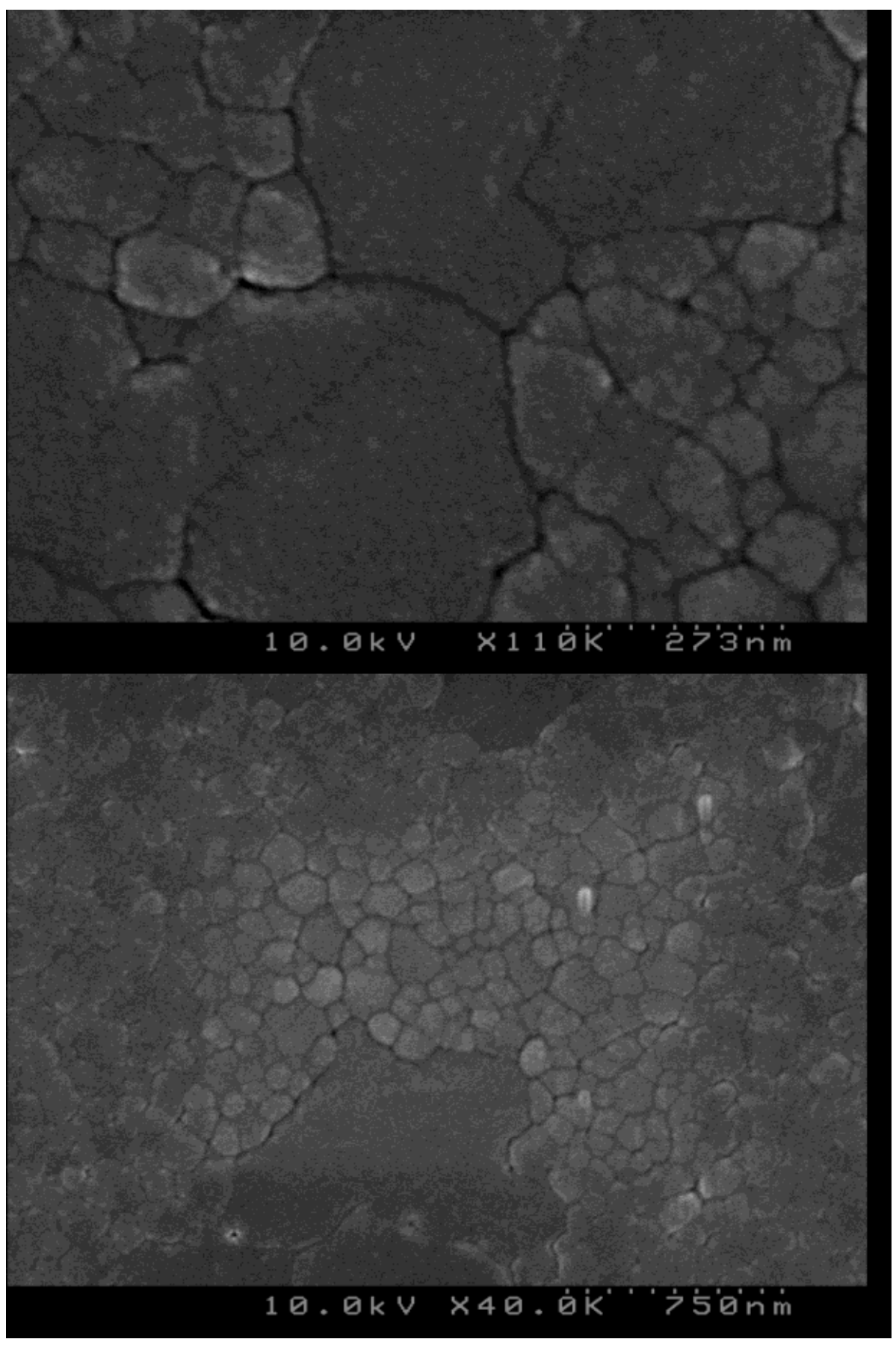

**Figure 7. SEM micrographs showing rutile grains alongside untransformed anatase in undoped films**

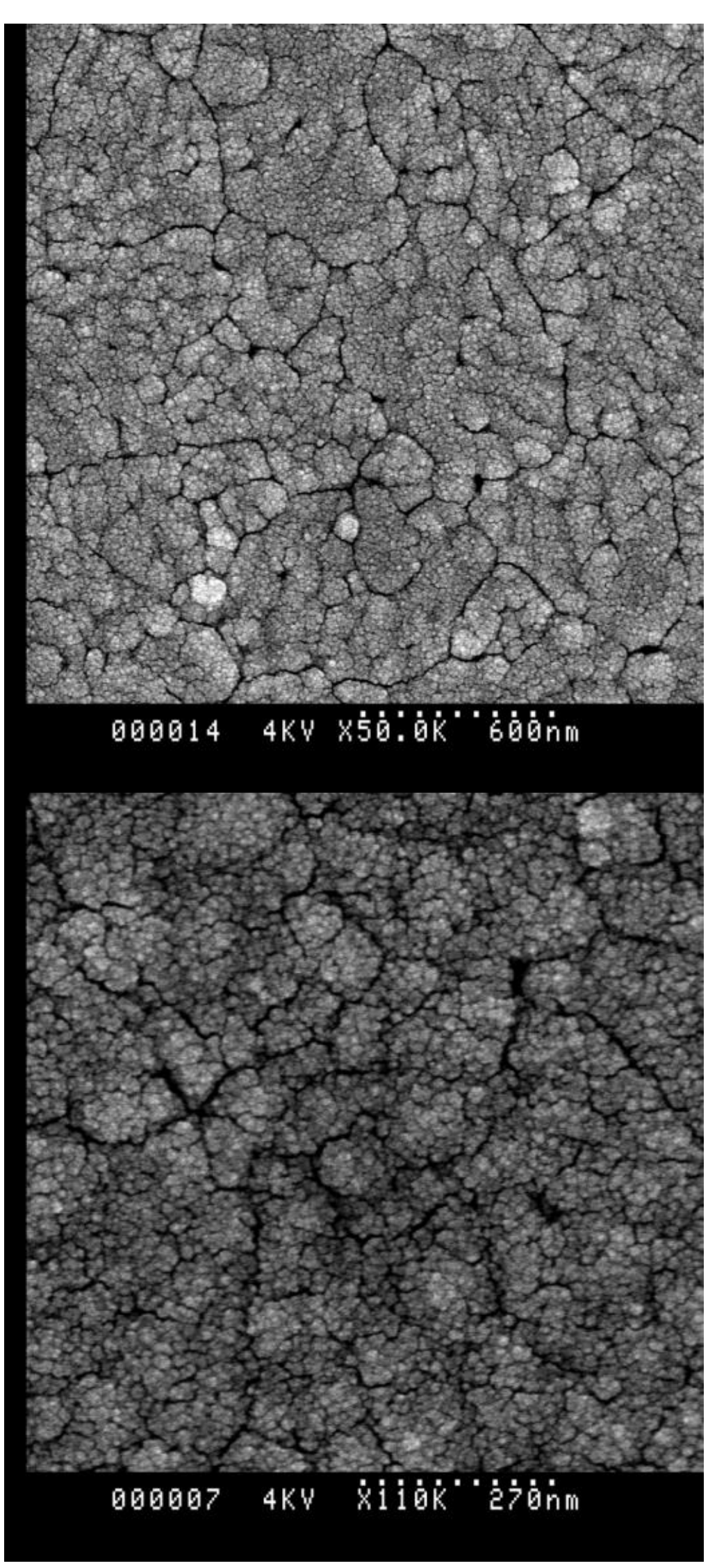

**Figure 8. (a) Cu/Fe co-doped mixed phase film (b) Cu doped rutile film**

## 3.5. UV-Vis Transmittance

UV-Vis transmittance measurements were carried out to ascertain the effects of dopants and phase assemblage on the optical properties of the thin films synthesised in this work. The transmittance of titania films was not used to make conjectures regarding the photonic efficiency as reduced transmittance does not infer superior optical absorbance as light may also be reflected or scattered. It can be seen that doped films show lower transmittance levels compared with undoped films. Films containing rutile phase, even at low phase proportions, showed an adsorption edge at higher wavelengths and lower overall transmittance in comparison with anatase films.

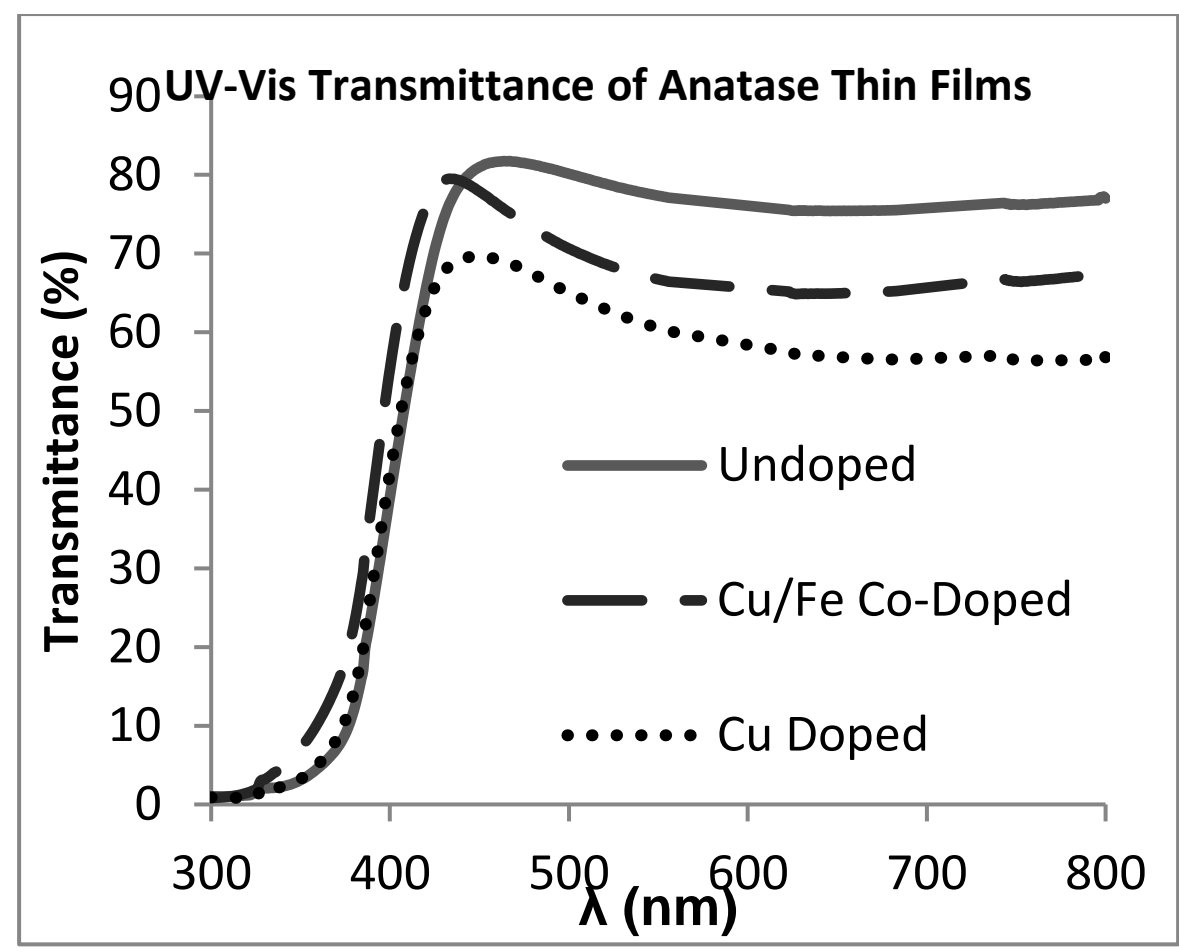

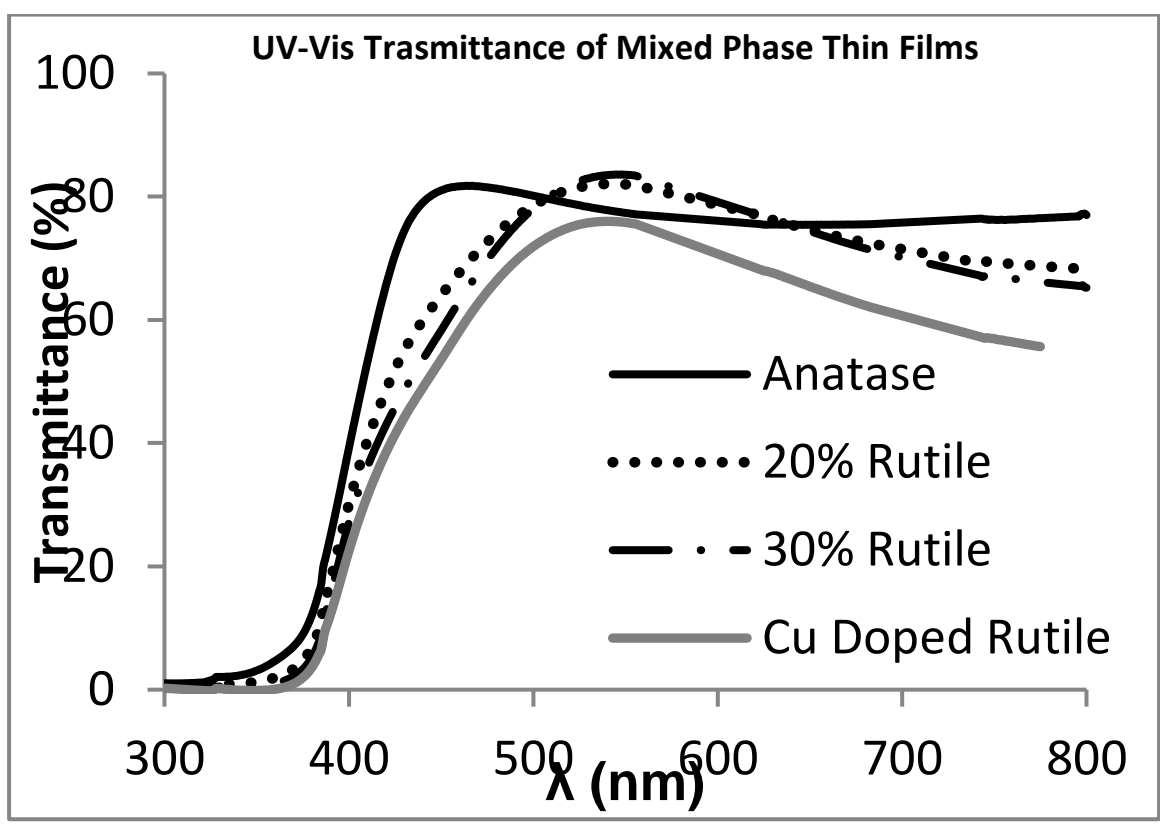

**Figure 9. (a) UV-Vis transmittance of anatase films (b)UV-Vis transmittance of mixed phase films**

## 3.6. Photocatalytic Activity

The photocatalytic activity of the thin films synthesised in this work was examined by the photo-degradation of methylene blue in aerated solutions. An uncoated quartz substrate was used to establish the baseline photo-degradation of the dye, which had an initial concentration of 0.107 mMol/Litre, in the absence of a photocatalyst. Figures 9(a) and 9(b) show the absorbance spectra of the solution subsequent to 10h and 20h of UV irradiation. These absorbance spectra were interpreted using the calibration curve shown in Figure 10 to establish the concentration of methylene blue in solution. The resultant dye concentrations are shown in Figure 11. It can be seen that the degradation of dye takes places to a significant extent without the presence of a $TiO_2$ film. While the photo-activity was poor in all films, it can be seen that the presence of $TiO_2$ enhances the degradation of the dye under UV light to a greater extent in undoped films than in doped films.

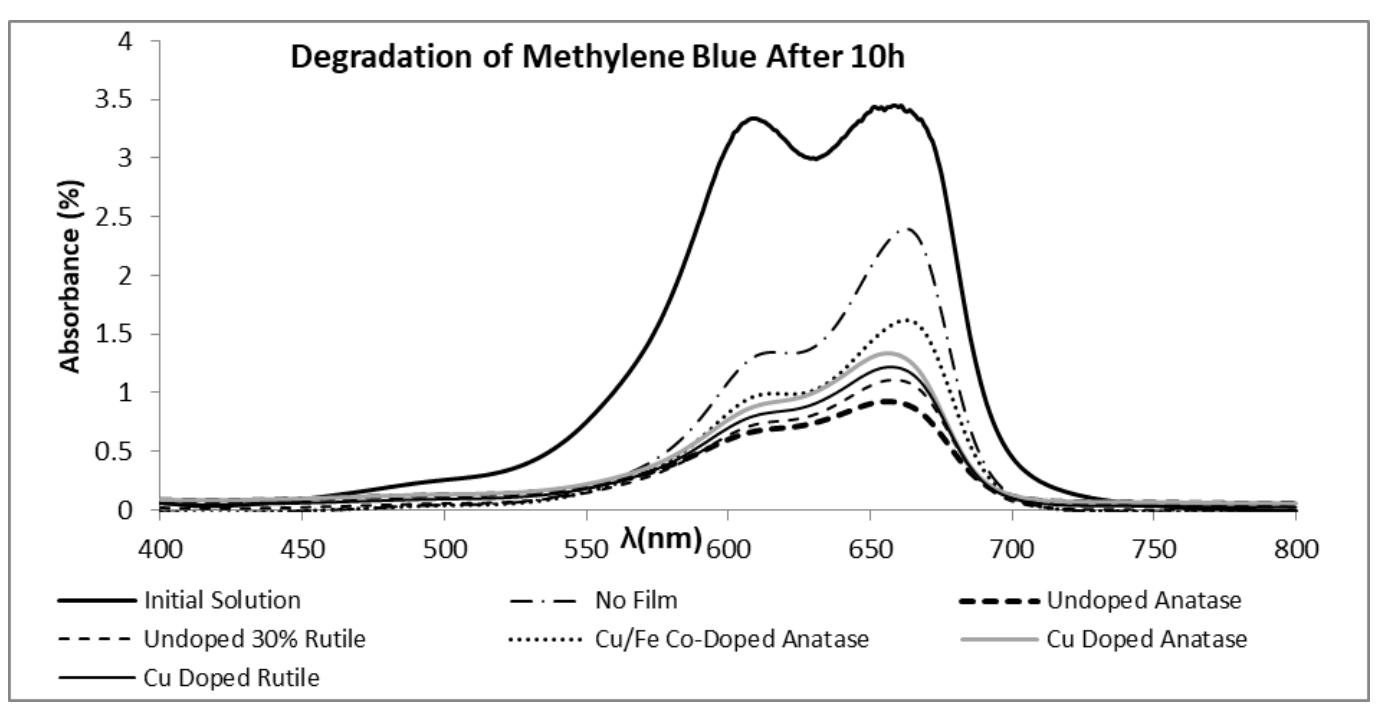

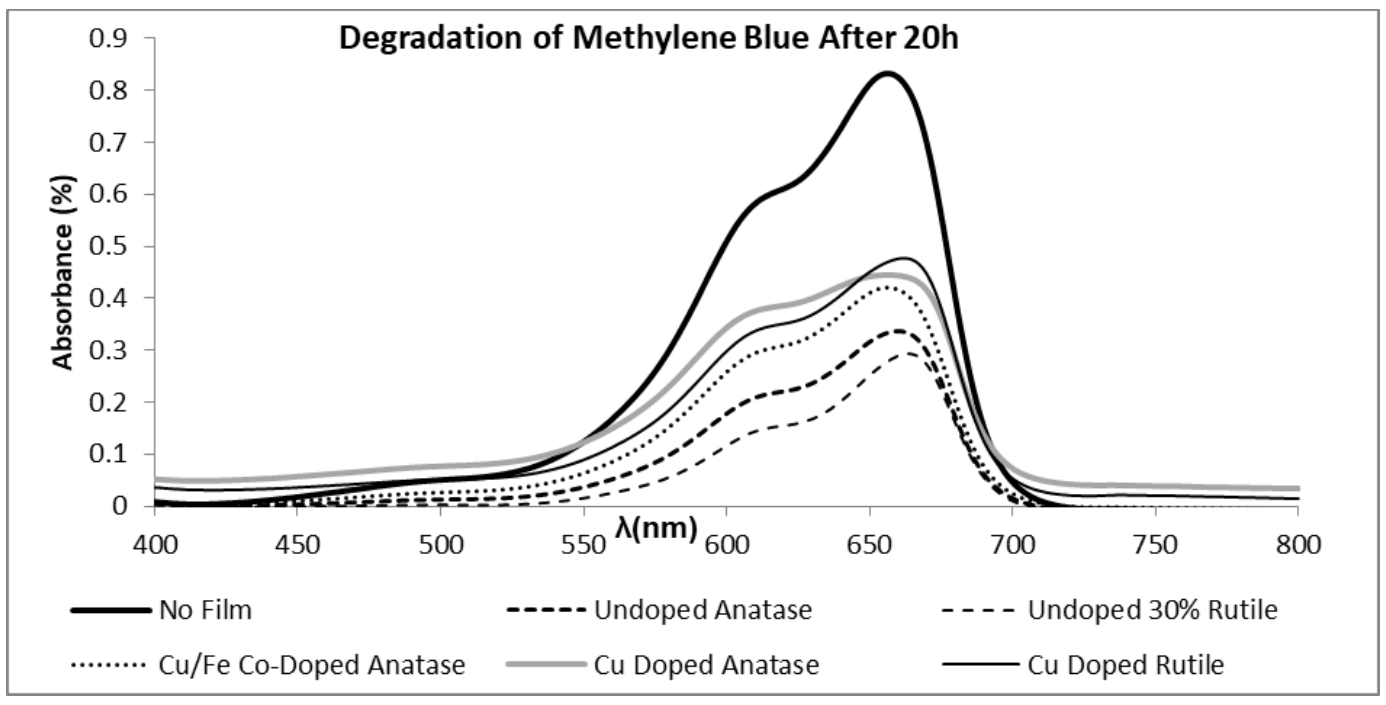

**Figure 10. Visible light absorbance of methylene blue solutions after (a) 10h and (b) 20h of UV**

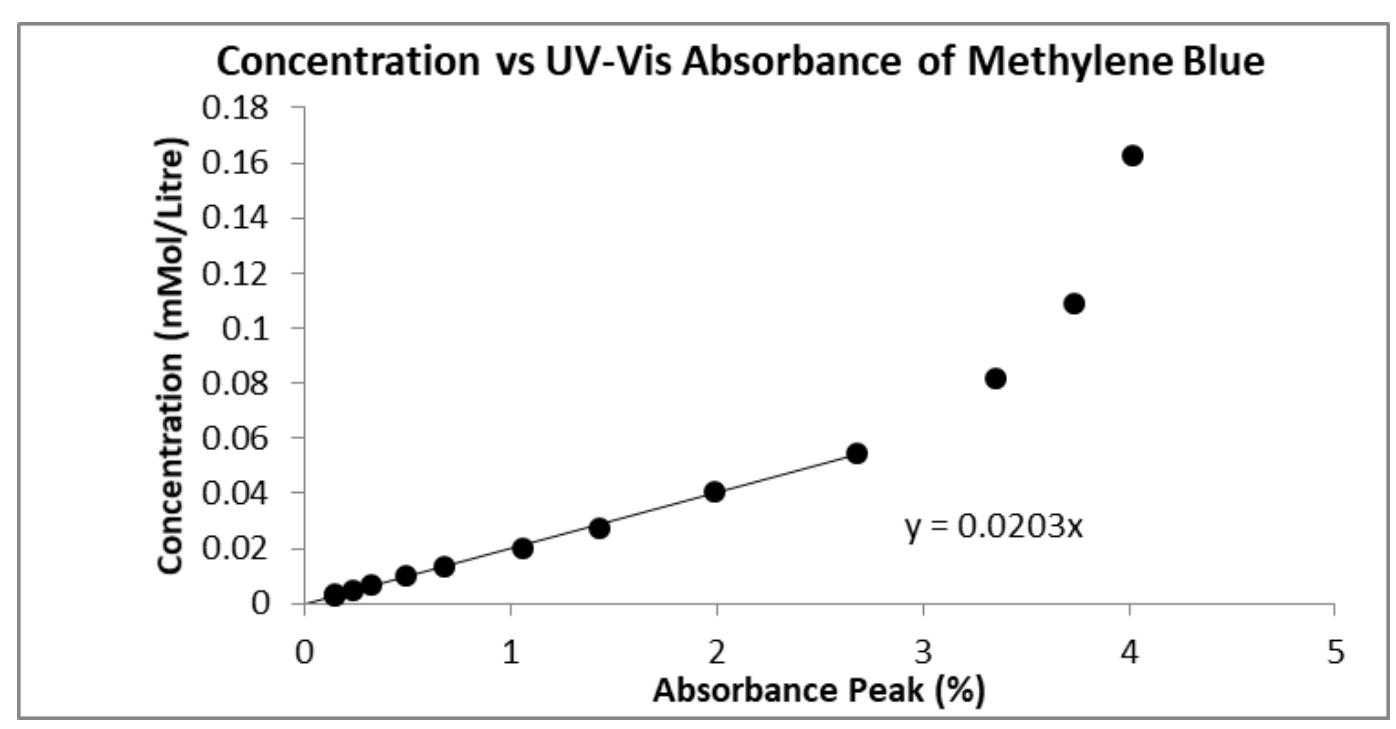

**Figure 11. Calibration curve used to determine dye concentration from optical absorbance data**

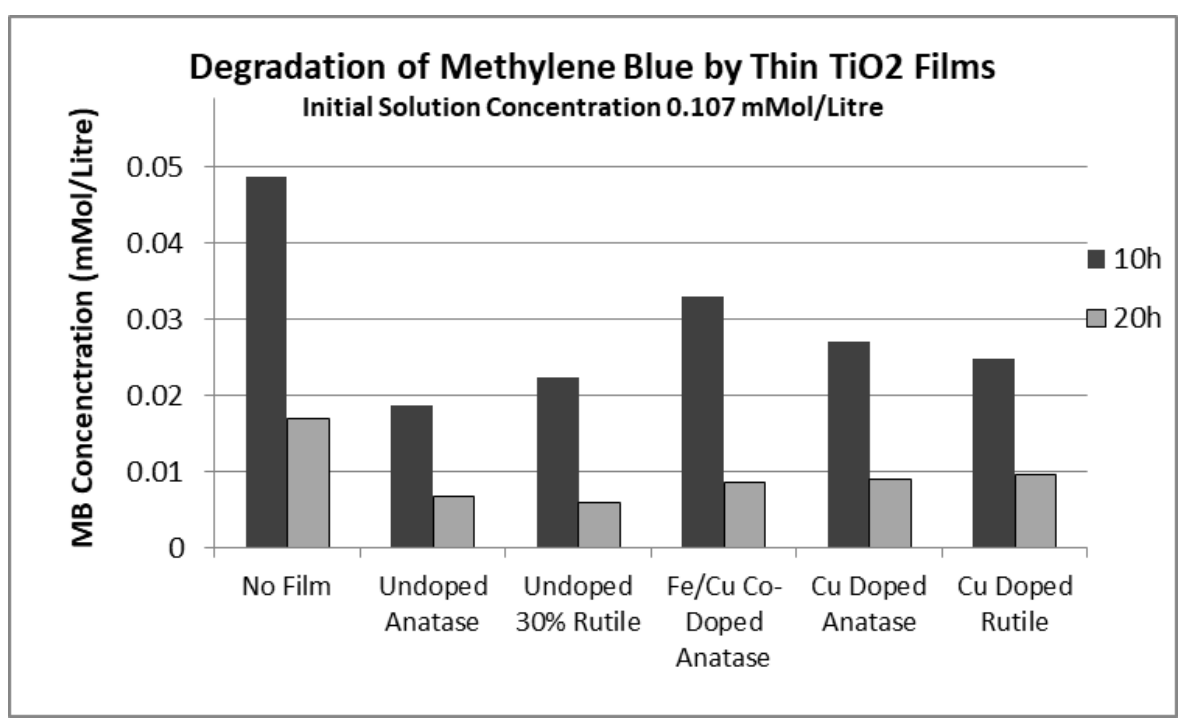

**Figure 12. $TiO_2$ photcatalyzed degradation of methylene blue**

# 4. Discussion

## 4.1. Phase transformation behaviour

From glancing angle XRD in can be seen that the anatase phase of $TiO_2$ films immobilised on quartz substrates shows improved thermal stability in comparison with unsupported $TiO_2$ powders. Undoped $TiO_2$ powder typically transforms to rutile at temperature of 600-700°C [39-41]. In contrast, undoped films synthesised in this work show the phase transformation onset only at 800°C. While it is possible that smaller temperature intervals and greater firing durations would reveal rutile formation at lower temperatures, the stabilisation of the anatase phase on quartz substrates due to diffusion of Si atoms is consistent with results published elsewhere [28, 42, 43]. The reconstructive anatase to rutile transformation involves the rearrangement of the $TiO_2$ lattice and an overall volume contraction of ~8% [44-46]. The presence of interstitial Si atoms is reported to hinder this structural rearrangement through lattice constraint [47] and it is further possible that the structural rearrangement is constrained further due to grain boundary pinning at the film-substrate interface.

In comparison with undoped $TiO_2$ thin films, the addition of Cu(ii) through the use of $CuCl_2$ additive to the precursor sol brought about a marked enhancement of the anatase to rutile transformation. The enhancement is evident from a more rapid transformation at lower temperatures as can be seen in Figure 3. This is likely to be a result of increased oxygen vacancies brought about by the substitution of Ti by lower valence Cu ions as reported elsewhere [29, 48, 49]. The presence of these oxygen vacancies reduces lattice constraint and facilitates the reconstructive phase transformation. The co-doping of films with Fe/Cu additives introduced using $CuCl_2$ and $FeCl_3$ results in a weak enhancement of the phase transformation possibly due to the larger $Fe^{3+}$ ions occupying interstitial lattice positions and impeding the structural rearrangement involved in the phase transformation.

## 4.2. Phase Morphology

In all XRD patterns the anatase (101) and rutile (110) peaks are clearly dominant with other peaks being difficult to detect, in part due to a high noise signal ratio. Although the determination of the Lotgering orientation factor was not possible due to the aforementioned high background signals, the dominance of the anatase (101) and rutile (110) peaks suggest films synthesised in this work are highly oriented. This is a natural consequence of very thin films being deposited on single crystal substrate and is similar to the high degree of orientation observed in thin $TiO_2$ films on single crystal substrates reported elsewhere [18, 50].

Raman microspectroscopy revealed that undoped $TiO_2$ films exhibited the formation of banded phase segregation with narrow parallel rutile bands widening to consume the parent anatase as the phase transformation progressed. The Parallel bands exhibited a single dominant orientation across the entire film area. The morphology of rutile formation observed in this work has not been reported in the literature to date.

In undoped titania, rutile has been reported to nucleate at twin interfaces and form as lathes in the parent anatase with rutile {110} planes parallel to anatase {112} planes [51-53]. Based on these reports rutile formation as parallel plates within anatase grains may be possible however in the case of phase transformation in films of isotropic anatase grains no overall dominant rutile would be expected. For this reason the banded appearance of undoped films suggests that the high degree of orientation in the film as observed by XRD patterns plays a role in governing the orientation of rutile growth. Such preferential orientations are likely to be the result of the interface with the single crystal quartz substrate. Consideration of film thickness and SEM micrographs suggests films synthesised in this work are likely to be of single grain thickness, particularly in the case of larger rutile grains, and this layer is likely to show significant substrate effect on crystallographic orientation.

The presence of continuous alternating bands of rutile and anatase as observed in undoped films was entirely absent from doped films. In doped films the rutile phase formed in segregated clusters or spots. This is likely to be due to the localisation of regions of increased oxygen vacancy levels owing to the segregation of anatase to rutile transformation enhancing dopants. As doped films were of similar thickness to undoped films we would expect similar substrate orientation effects as seen in undoped films, however anisotropic rutile growth was not observed in such films. This suggest localised effects of dopants outweigh crystallographic orientation effects in influencing the phase transformation behaviour in thin films.

## 4.3. Film Morphology

SEM analysis of mixed phase undoped films revealed the presence of large consolidated grains alongside smaller grains. This is likely to be the result of the significant grain coarsening which takes place as anatase transforms to rutile. In both phases little or no porosity is evident. The coarsening is significant as anatase grains of ~40nm transform to rutile of 300-400 nm size. This grain coarsening is consistent with reports in the literature and is generally considered detrimental to photocatalytic performance of titania as larger grains exhibit lower surface area [18, 48, 54, 55]. The doped films synthesised in this work appeared to show less grain growth as mixed phase doped films did not exhibit larger consolidated grains as was the case with undoped films, reports elsewhere have also observed that dopants can facilitate the anatase to rutile phase transformation without significant grain growth[55]. As films in this work are fully dense, the effect of grain growth on effective surface and adsorption of reactive species to the catalyst substrate is likely to be less significant in comparison with the use of powders in suspension or porous materials.

## 4.4. Optical Properties

UV-Vis transmittance cannot be used to evaluate the photonic efficiency of photocatalytic films as light that is not transmitted may be scattered or reflected. Furthermore an increase in optical absorbance does not necessarily infer an increase in exciton photo-generation. Thus it would be erroneous to make any assumptions regarding the merit of films as photocatalysts based on the measurement of optical properties.

It appears that dopants bring about a reduction in optical transmittance. This effect is likely to be due to colour effects of the transition metal dopants used. These effects would be greater however the film thickness is less than the wavelength of the light used in UV-Vis transmittance measurements. Co-doped films did not exhibit a notable change in optical properties suggesting that intervalence charge transfer between $Cu^{2+}$ and $Fe^{3+}$ does not take place to a significant extent.

The presence of rutile in the films altered the optical transmittance of the films shifting the absorption edge to higher wavelengths (lower photon energies). This change in optical absorption is a result of the lower band gap of rutile $TiO_2$. The presence of even a small level of rutile phase is sufficient to significantly alter the UV-Vis transmittance, this is evident in the similarities between the transmittance of a 20% rutile film and a 100% rutile film.

## 4.5. Photocatalytic Performance

As shown by the results from the control sample (a quartz substrate without a $TiO_2$ film), methylene blue concentrations are reduced by exposure to UV in the absence of a photocatalyst. The presence of $TiO_2$ only brings about a minor reduction in dye concentration in comparison with uncoated quartz suggesting poor photocatalytic activity. This low photocatalytic activity could be due to low $TiO_2$ surface area which results from the high density of the film and lack of dye molecule adsorption to the film surface.

Photocatalytic photo-degradation of methylene blue was not noticably enhanced in bi-phasic films and a comparison of the dye photo-degredation by single phase and mixed phase films does not show a significant variation. The presence of dopants appears to have negative effects on the performance of thin film photocatalysts this is likely to be due to dopants acting as charge carrier recombination centres as reported elsewhere. As methylene blue decomposition experiments has not yielded conclusive results, it is possible that photo-luminescence would enable further investigation of exciton lifetime.

## 5. Conclusions

- Sol-Gel Synthesis of $TiO_2$ thin films on single crystal quartz substrates yielded highly oriented and thermally stable anatase.
- The addition of Cu enhances the transformation of anatase to rutile through an increase in oxygen vacancies in the $TiO_2$ lattice
- Undoped thin films show the formation of rutile as oriented parallel bands.
- Transition metal dopants segregate in $TiO_2$ thin films and bring about a localised isotropic formation of the rutile phase.
- The presence of low levels of rutile is sufficient to alter the optical transmittance of $TiO_2$ thin films
- Transition metal doping and co-doping is detrimental to photocatalytic performance

**Acknowledgements**

The authors acknowledge access to the UNSW node of the Australian Microscopy and Microanalysis Research Facility (AMMRF) and the spectroscopy facilities at the Mark Wainwright Analytical Centre.